\documentclass{ws-procs975x65}

\begin{document}

\title{Is Torsion a Fundamental Physical Field?}

\author{O. M. LECIAN$^{12a}$, S. MERCURI$^{12b}$, G. MONTANI$^{123c}$}

\address{$^{1}$ICRA --- International Center for Relativistic Astrophysics.\\
$^{2}$Dipartimento di Fisica, Universit\`a di Roma ``La Sapienza'', P.le
Aldo Moro 5,\\
00185 Roma, Italy.\\
$^{3}$ENEA C.R. Frascati (Dipartimento F.P.N.), Via Enrico Fermi 45,\\
00044 Frascati, Roma, Italy.\\
$^{a}$E-mail: lecian@icra.it
$^{b}$E-mail: mercuri@icra.it
$^{c}$E-mail: montani@icra.it}

\begin{abstract}
The local Lorentz group is introduced in flat space-time, where the resulting Dirac and Yang-Mills equations are found, and then generalized to curved space-time: if matter is neglected, the Lorentz connection is identified with the contortion field, while, if matter is taken into account, both the Lorentz connection and the spinor axial current are illustrated to contribute to the torsion of space-time. 
\end{abstract}


\bodymatter

\paragraph{Lorentz gauge theory on flat space-time}\hfill\\
Let $M^{4}$ be a 4-dimensional flat manifold equipped with the metric tensor $g_{\mu\nu}=\eta_{\alpha\beta}\frac{\partial x^{\alpha}}{\partial y^{\mu}}\frac{\partial x^{\beta}}{\partial y^{\nu}}\equiv\eta_{\alpha\beta}e^{\alpha}_{\mu}e^{\beta}_{\nu}$, where $e^{\alpha}_{\mu}$ are bein vectors, $x^{\alpha}$ are Minkowskian coordinates, and $y^{\mu}$ are generalized coordinates. Under an infinitesimal generic diffeomorphism, $x^{\alpha}\rightarrow y'^{\mu}=\delta^{\mu}_{\alpha}x^{\alpha}+\xi^{\mu}(x^{\gamma})$,
and for an infinitesimal local Lorentz transformation ,$x^{\alpha}\rightarrow x'^{\alpha}=x^{\alpha} +\epsilon^{\alpha}_{\ \beta}(x^{\gamma})x^{\beta}$, the behavior of a vector field $V_{\alpha}\rightarrow V'_{\alpha}$ must be equivalent: from the comparison of the two transformation laws, the identification $\epsilon _{\alpha}^{\ \beta}\equiv \frac{\partial \xi_{\alpha}(x^{\gamma})}{\partial x^{\beta}}$ is possible, where the isometry condition $\partial_{\beta}\xi_{\alpha}+\partial_{\alpha}\xi_{\beta}=0$ has been taken into account.\small \\
Spinor fields, on the contrary, cannot have the same behavior under the two transformations, for spinors transform under a spinor representation of the Lorentz group, while no spinor representation is given for the diffeomorphism group \cite{hehvdhker76}, i.e. spinor fields must experience the isometric component of the diffeomorphism as a local Lorentz transformation. The implementation of a local symmetry requires the introduction of gauge field, and the space where these gauge transformations live can be defined by comparing the coordinate transformation that induces vanishing Christoffel symbols in the point $P$, $\ y^{\alpha}_{P}=x^{\alpha}_{tb}+\frac{1}{2}\left[\Gamma^{\alpha}_{\beta\delta}\right]_{P}x^{\beta}_{tb}x^{\delta}_{tb}$,
where $tb$ refers to the tangent bundle,
with the generic diffeomorphism $y^{\alpha}=x^{\alpha}+\xi^{\alpha}(x^{\gamma})$: 
the identification, in the point $P$,$x^{\alpha}_{P}=x^{\alpha}_{tb}+\frac{1}{2}\left[\Gamma^{\alpha}_{\beta\delta}\right]_{P}x^{\beta}_{tb}x^{\delta}_{tb}-\xi^{\alpha}$ is possible \cite{mm}. The coordinates of the tangent bundle are linked point by point to those of the Minkowskian space through the relation above, and they differ for the presence of the infinitesimal displacement $\xi$. From now on, these coordinates will be referred to as $x^{a}$.\\
Let $M^{4}$ be a 4-dimensional flat Minkowski space-time: the action describing the dynamics of spin-$\frac{1}{2}$ fields, $S=\frac{i}{2}\int d^4x\left(\overline{\psi}\gamma^{a}\partial_{a}\psi-\partial_{a}\overline{\psi}\gamma^{a}\psi\right)$, is invariant under global Lorentz transformations $\psi\rightarrow S\left(\Lambda\right)\psi, \overline{\psi}\rightarrow
\overline{\psi}S^{-1}\left(\Lambda\right)$, where $S\left(\Lambda\right)=1-\frac{i}{4}\epsilon^{ab}\Sigma_{ab}$ is the infinitesimal global Lorentz transformation, defined as in \cite{ms}.\\ \small
For local Lorentz transformations, the Lagrangian density will read $L=\frac{i}{2}\,e^{\mu}_{\phantom1a}\left[\overline{\psi}\gamma^{a}D_{\mu}\psi-D_{\mu}\overline{\psi}\gamma^a\psi\right],\ \ D_{a}\psi=e^{\mu}_{\phantom1a}D_{\mu}\psi=e^{\mu}_{\phantom1a}\left(\partial_{\mu}\psi-\frac{i}{4}A_{\mu}^{b c}\Sigma_{b c}\psi\right)$ being the pertinent covariant derivatives. The interaction Lagrangian density $\mathcal{L}_{int}=\frac{1}{8}\,e^{\mu} _{\phantom1a}\,\overline{\psi}\left\{\gamma^{a},\Sigma_{b c}\right\}\psi A^{b c}_{\mu}=-S_{b c}^{\mu}A^{b c}_{\mu}=-\frac{1}{4}\,\epsilon^{a b}_{\phantom1\phantom1c d}e_{\mu}^{\phantom1c}j_{A}^{d}A^{b c}_{\mu}$,
where $j_{A}^d=\overline{\psi}\gamma_5\gamma^d\psi$ is the spinor axial current, shows that the gauge field $A$ interacts with the spinor axial current, which is the source for gauge field of the Lorentz group on flat space-time. After variation with respect to the adjoint field, and making use of the anti-commutation properties of Dirac matrices, the Dirac equation $e^{\mu}_{\phantom1a}\left[i\gamma^{a}\partial_{\mu}+\frac{1}{4}\,\epsilon^{a b}_{\phantom1\phantom1c d}\gamma_5\gamma_b A^{c d}_{\mu}\right]\psi=0$ for the spinor $\psi$ in an accelerated frame is found: the spinor cannot be considered as free, because it interacts with a Yang-Mills gauge field. If a Lagrangian density for the gauge field $A$ is added, i.e. $L=-(1/32)tr\,\star F\wedge F$, variation with respect to $A$ leads to the Yang-Mill equation $D_{\mu}F^{\mu\nu a}_{\ \ \ \ b}=S^{\nu a}_{\ \ \ b}$. \small\\

\paragraph{Lorentz gauge theory on curved space-time}\hfill\\
The need to introduce a Lorentz gauge field in curved space-time comes from the fact that, while spin connections are intended to restore the properties of Dirac matrices in the physical space-time, gauge connections allow one to recover invariance under local Lorentz transformations for spinor fields on the tangent bundle\footnote{On curved space-time two different Lorentz transformations can be distinguished, which coincide in flat space-time.
Active Lorentz transformations are due to the action of the Lorentz group on tensors $V^{\mu}$ and spinors $\psi$ on the tangent bundle, i.e. $V^{\mu}\rightarrow \Lambda(x)^{\mu}_{\ \nu}V^{\nu}$ and $\psi\rightarrow s(\Lambda(x))\psi$.Passive Lorentz transformations are due to isometric diffeomorphisms of the space-time manifold, which pull back the local basis in the generic point $P$. While active transformations are defined everywhere once the matrix $\Lambda (x)$ is assigned, passive transformations can be reduced to a Local Lorentz transformation only in the point $P$, acting as a pure diffeomorphism on the other points of the manifold. These two kinds of transformations, indeed, coincide on curved space-time, too: because of local Lorentz transformations, a tetradic vector transforms as $e'^{\bar{a}}_{\mu}(x')= \Lambda^{\bar{a}}_{\ \bar{b}}(x')e^{\bar{a}}_{\mu}(x')$, while, for world transformations, $e^{\bar{a}}_{\mu}(x)\rightarrow e'^{\bar{a}}_{\mu}(x')=e^{\bar{a}}_{\mu}(x)\frac{\partial x^{\rho}}{\partial x'^{\mu}}\approx e^{\bar{a}}_{\mu}(x)+e^{\bar{a}}_{\rho}(x)\frac{\partial \xi^{\rho}}{\partial x'^{\mu}}$. The comparison leads to the identification $e'^{\bar{a}}_{\mu}(x')=
e^{\bar{a}}_{\mu}(x')+e^{\bar{b}}_{\mu}(x')\epsilon^{\bar{a}}_{\bar{b}}$, where $\epsilon^{\bar{a}}_{\ \bar{b}}\equiv  -D_{\bar{b}}\xi^{\bar{a}}-R^{\bar{a}}_{\ \bar{b}\bar{c}}\xi^{\bar{c}}$, $\lambda_{\bar{a}\bar{b}\bar{c}}=R_{\bar{a}\bar{b}\bar{c}}-R_{\bar{a}\bar{c}\bar{b}}$ being the anolonomy coefficients.\small}.(For a first attempt to a gauge theory of the gravitational field, see \cite{uti56}\cite{kib61}).\\
As a consequence, two different Lorentz-valued 1-forms are required to make the spinor Lagrangian density invariant: the total connection reads $C^a_{\phantom1b}=\omega^a_{\phantom1b}+A^a_{\phantom1b}$, where $\omega$ is the usual spin connection of GR, and $A$ is an additional Lorentz connection and the total action reads\footnote{The interaction term between $w$ and $A$ is added by hand, and will be crucial for the geometrical interpretation of the Lorentz-group field. We are assuming $8\pi G=1$.}
\begin{align}\label{total action}
\nonumber S&\left(e,\omega,A,\psi,\overline{\psi}\right)=\\
  &\frac{1}{4}\int\epsilon_{a b c d}\,e^{a}\wedge e^{b}\wedge R^{c d}-\frac{1}{32}\int tr\,\star F\wedge F-\frac{1}{4}\int\epsilon_{a b c d}\,e^{a}\wedge e^{b}\wedge\omega^{[c}_{\phantom1f}\wedge A^{f d]}+\nonumber
\\
& \frac{1}{2}\int\epsilon_{a b c d}\,e^{a}\wedge e^{b}\wedge e^{c}\wedge\left[i\overline{\psi}\gamma^d\left(d-\frac{i}{4}\left(\omega+A\right)\right)\psi-i\left(d+\frac{i}{4}\left(\omega+A\right)\right)\overline{\psi}\gamma^d\psi\right].
\end{align}

If fermion matter is absent, variation with respect to the connection gives the structure equation $d^{(\omega)}e^a=A^a_{\phantom1b}\wedge e^b$: pulling back the action to the unique solution\footnote{For a discussion of the reduction of the dynamics, see \cite{art}.}  $\omega^a_{\phantom1b}=\widetilde{\omega}^a_{\phantom1b}+A^a_{\phantom1b}$, we get \small
\begin{align}\label{reduced action}
\nonumber S&\left(e,A\right)=\frac{1}{4}\int\epsilon_{a b c d}\,e^{a}\wedge e^{b}\wedge\widetilde{R}^{c d}-\frac{1}{32}\int tr\,\star F\wedge F+\\
&-\frac{1}{4}\int\epsilon_{a b c d}\,e^{a}\wedge e^{b}\wedge\widetilde{\omega}^{[c}_{\phantom1f}\wedge A^{f d]}-\frac{1}{4}\int\epsilon_{a b c d}\,e^{a}\wedge e^{b}\wedge A^{c}_{\phantom1f}\wedge A^{f d},
\end{align}
where $tilde$ denotes Riemannian objects. Variation with respect to the gravitational field and to the Lorentz connection gives \small
\begin{align}\label{Einstein total equations}
\epsilon^a_{\phantom1b c d}\,e^{b}\wedge\widetilde{R}^{c d} & =M^a+\epsilon^a_{\phantom1b c d}\,e^{b}\wedge\left(\widetilde{\omega}^{c}_{\phantom1f}+A^{c}_{\phantom1f}\right)\wedge A^{f d},
\\
d^{(A)}\star F^{f d} & =\epsilon_{a b c}^{\phantom1\phantom1\phantom1[d}\,e^a\wedge e^b\wedge\left(\omega^{c f]}+2A^{c f]}\right),
\end{align}
where $M^a$ is the energy-momentum 3-form of the field $A$, which can be explicitly obtained variating the Yang-Mills- like action with respect the gravitational 1-form. Since the solution to the structure equation is analogous to that of the $2^{nd}$ Cartan structure equation, the Lorentz connection $A$ can be identified with the contortion 1-form, thus implying the presence of the torsion 2-form $T^a=A^a_{\phantom1b}\wedge e^b$. Field equations describe the coupling between gravitational and Lorentz connections: gravitational spin connections become the source of torsion.\\    
\small
If fermion matter is present, variation with respect to the connections give the structure equation
$d^{(\omega)}e^a=A^a_{\phantom1b}\wedge e^b-\frac{1}{4}\epsilon^a_{\phantom1b c d}e^b\wedge e^c j_{(A)}^{d}$: pulling back the action to its unique solution   $\omega^a_{\phantom1b}=\widetilde{\omega}^a_{\phantom1b}+A^a_{\phantom1b}+\frac{1}{4}\epsilon^a_{\phantom1b c d}e^c j_{(A)}^{d}$, we obtain
\begin{align}\label{total action reduced}
\nonumber S&\left(e,A,\psi,\overline{\psi}\right) =\frac{1}{2}\int\epsilon_{a b c d}\,e^{a}\wedge e^{b}\wedge \widetilde{R}^{c d}-\frac{1}{32}\int tr\,\star F\wedge F+
\\\nonumber
& +\frac{1}{2\cdot 3!}\int \epsilon_{a b c d}\,e^{a}\wedge e^{b}\wedge e^{c}\wedge\left[i\overline{\psi}\gamma^d\left(d-\frac{i}{4}\left(\widetilde{\omega}+A\right)\right)\psi-i\left(d+\frac{i}{4}\left(\omega+A\right)\right)\overline{\psi}\gamma^d\psi\right]+
\\\nonumber
& -\int\epsilon_{a b c d}\,e^{a}\wedge e^{b}\wedge A^{c}_{\phantom1f}\wedge A^{f d}-\int\epsilon_{a b c d}\,e^{a}\wedge e^{b}\wedge\widetilde{\omega}^{[c}_{\phantom1f}\wedge A^{f d]}+
\\
& -\frac{3}{16}\int e_a\wedge e_b\wedge e_c \wedge A^{[a b}\,j^{c]}_{(A)}-\frac{3}{16}\int d^4x\,\eta_{a b}j_{(A)}^a j_{(A)}^b,
\end{align}
where the last term is the four-fermion interacting term of Einstein-Cartan theory. The presence of spinor fields in the structure equation means that both the connection $A$ and the spinor axial current contribute to the torsion of space-time. Variation with respect to the gravitational field and to the Lorentz connection leads to the generalization of the field equation obtained in absence of matter.

\end{document}